\documentstyle[preprint,aps,epsf]{revtex}
\newcommand{\cleb}[6]{C^{#1#2#3}_{#4#5#6}}
\newcommand{\lam}{\lambda}
\newcommand{\beq}{\begin{equation}}
\newcommand{\eeq}{\end{equation}}
\newcommand{\ba}{\begin{array}}
\newcommand{\ea}{\end{array}}
\newcommand{\beqa}{\begin{eqnarray}}
\newcommand{\eeqa}{\end{eqnarray}}
\newcommand{\bd}[1]{ \mbox{\boldmath $#1$}  }

\begin{document}
\draft

\title{Long Range Alpha Characteristics in the 
Ternary Cold Fission of $^{252}$Cf}
\author{\c Serban Mi\c sicu, A.Sandulescu, F.Carstoiu, M.Rizea and W.Greiner}
\address{Department for Theoretical Physics, 
National Institute for Nuclear Physics, Bucharest-M\u agurele, ROMANIA}
\maketitle
\begin{abstract}
We compute the final kinetic energies of the fragments 
emitted in the light charged particle accompanied cold fission 
of $^{252}$Cf  taking into account the deformation and the finite-size 
effects of the fragments and integrating the equations of motion 
for a three-body system subjected only to Coulomb forces. 
The initial conditions for the trajectory calculations were derived 
in the frame of a deformed cluster model which includes also the effect
due to the absorbative nuclear part. Although the distributions of initial 
kinetic energies is rather broad we show that in cold fission 
the initial conditions can be better determined than in the usual
spontaneous fission. 
\end{abstract}

\newpage

\section{Introduction}

The cold ternary fission is a rearrangement process of a large group of 
nucleons from the ground state of the initial nucleus to the ground state
of the three final fragments. Like in the case of spontaneous and 
thermal-induced fission a ternary component of a few tenths of percent is 
present also in the cold fission process \cite{Wag91,Goen89,Flor2}.

In order to determine the configuration and the dynamics of the 
fissioning nucleus at scission, the experimental data for the light 
charged particle (LCP) emitted in the fragmentation process are analyzed 
and compared with the theoretical results obtained via trajectory 
calculations. In the past a large number of studies were devoted to the 
trajectory calculation, specially for $\alpha$-particles in the point 
charge approximation and without the account of nuclear forces. The 
alphas were considered to be emitted from the neck region   
\cite{GH65,BFN67,Halp71,Ert69,Fo70,CR78}. There have been also some
authors who considered the finite size and the deformation effects  
\cite{CL80,RRD82,PP84} and showed that these geometrical factors are 
influencing sensitively the angular distributions of the LCP.

However in all these approaches to the ternary spontaneous fission the 
problem of choosing the initial parameters of the trajectory calculations 
is complicated by the fact that various theories give different 
predictions. Since only the energies and the angles of the three 
particles can be experimentally determined, solving the equations of 
motion backward in time will not provide a full information on the 
geometric and dynamic characteristics of the fissioning system at the
moment of the LCP-emission. The only possibility is to probe various 
combinations of assumed initial conditions and then compute the 
trajectories for comparison with the available experimental data. In the 
{\em hot} ternary fission the initial conditions are so numerous that in 
order to encompass as much as possible combinations, Monte-Carlo 
techniques were employed \cite{RRD82,PP84}.

For the cold ternary fission the initial conditions are better known
\cite{Flor2,SCM98}. First of all the fragment deformations are those of 
their ground states. This fact prompted us to calculate the final 
characteristics of the LCP emitted in the cold fission  of $^{252}$Cf 
for different mass splittings, and see how the 
static deformations and the finite size are modifying the outcome of 
the trajectory calculation. Using forces, computed through  
a double folded integration of the Coulomb interaction between two 
quadrupole deformed heavy ions, we derived the equations of motion for 
the three-body problem, and solved them numerically . 
The solution of this set of equations provided
the final angle of the LCP with respect to the fission axis and its 
kinetic energy.        
We compared our calculations with the point-like trajectory and some 
experimental consequences were discussed. 

\section{Determination of the initial conditions}

Usually, in trajectory calculations for the spontaneous fission 
different choices are taken for the initial kinetic energies of
the fragments emitted in the process. For example the initial kinetic energy 
of the two main fragments and of the $\alpha$ emitted in the spontaneous 
ternary fission should be around 0.5 MeV according to the statistical 
theory and  the equipartition principle\cite{Ert69,Fo70}.
On the contrary, in the dynamical theory of fission \cite{NS76} the nascent 
fragments at scission are predicted to be moving with appreciable kinetic
energy (20-50 MeV).  
The initial velocities of the heavy fragments are considered to have non-zero 
components only along the $x$-axis. The initial velocity of the light fragment 
$v_L(0)$ is related to the initial velocity of the heavy fragment $v_H(0)$
in such a way that the total momentum of the two fission fragments is zero 
along the $x$-axis, i.e. $v_L(0) = {A_H\over A_L} v_H(0)$. This reasonable 
assumption will be applied also by us.

Therefore we have to determine the following initial conditions :
a) The tip distance $d$; b) the kinetic energies of the two main fragments
$E_{H0}$, $E_{L0}$ and the kinetic energy of the LCP $E_{LCP}$; c) the initial 
geometric configuration of the LCP, i.e. the  position between the fragments 
and the angle $\theta_{LCP}^0$ between the direction of motion of the LCP and 
the axis joining the two main fragments; d) The shape of the 
fragments. 

The determination of these quantities in the ternary cold fission will be 
facilitated, up to a certain extent, by the peculiar characteristic of the 
process, i.e. the fragments are emitted with total kinetic energy $TKE$ 
close to the corresponding ternary decay energy $Q_t$. In order to achieve 
such large $TKE$ values, the three final fragments should have very compact 
shapes at the scission point and deformations close to those of their 
ground states, similar 
to the case of the cold binary fragmentations. One may next suppose that the 
shapes of the fragments will not be modified when the fragments move away in 
the Coulomb field of each other. Thus, the problem of the fragments shape in 
the initial configuration is easily established for the cold fission.   

In order to determine the kinetic energies of the two main fragments we make 
use of the considerations derived from the deformed cluster model that we 
employed in previous papers for the study of the ternary cold fission 
\cite{SCM98}. In this deformed cluster model the barrier between heavy 
fragments (for binary fission) and the barrier between the LCP and the 
heavier fragments (for ternary fission) can be calculated quite 
accurately due to the fact that the touching configurations are situated 
inside of the barriers. For the two fragments, the exit point from the 
potential barrier is at a tip distance $d$ around 3 fm, as can be seen 
in Figure 1, for the case $^{248}$Cm $\rightarrow$ $^{104}$Mo + $^{144}$Xe.
This barrier is much thiner than the barrier between the LCP and the heavier 
fragments, and thus in our model first the two heavier fragments penetrate 
the potential barrier between them and later on the LCP is emitted.
Consequently the mass distributions of the heavier fragments are very similar
to those of the cold binary fission of an initial nucleus leading to the same
heavy fragments, i.e. $^{248}$Cm if the LCP is an $\alpha$ \cite{SFCG96}, or 
$^{242}$Pu if the LCP is $^{10}$Be \cite{SCM98}. The decay energy for such a 
binary fragmentation will be $Q_{LH}=Q_t - Q_{LCP}$, where $Q_t$ is the 
ternary decay energy of $^{252}$Cf and $Q_{LCP}$ is 6.22 MeV for $\alpha$ and 
8.71 MeV for $^{10}$Be. 
  
On ground of the cold fission characteristics mentioned above one may 
conjecture that at the exit point (second turning point) of the two 
heavier fragments, their potential energy is equal to $Q_{LH}$ and their 
kinetic energy is equal to zero. When the fragments move apart, i.e. their tip 
distance increase, their kinetic energy increase to.
In order to estimate the total kinetic energy of the fragments we have to 
find out at which tip distance the release of the LCP is likely to occur and 
compute at that point the potential energy, i.e.
\beq
TKE(d)\equiv TKE_L + TKE_H =Q_{LH}-V_{LH}(d)
\eeq
Using the conservation of linear momentum invoked above we have
\beq
TKE_L = {A_H\over A_L} TKE_H
\eeq
and the individual kinetic energies in terms of the total
kinetic energy read
\beq
TKE_i = {A_i\over A_H+A_L}TKE(d)~~~~(i=L,H)
\eeq
Now we turn to the problem of determining the tip distance $d$. It is 
reasonable to suppose that $d$ should correspond to the configuration at 
which the LCP is released. In Figures 2a and 2b we plotted the ternary 
potential seen by the LCP (in this case an $\alpha$) in the field of the 
two heavy fragments. As we shall se later the LCP should stay between 
the two heavy fragments in a position which should avoid its absorbtion 
by any of the fragments. We see in Figure 2a, that for tip distances 
up to 7 fm, the $\alpha$ is facing a thick barrier in the transversal 
direction. Eventually as the distance between the fragments increases the 
pocket in which the $\alpha$ is located becomes more and more shallower
untill it disapears around $d$ = 8 fm. Therefore one may conclude 
from these qualitative arguments that the initial tip distance 
between the two main fragments should not be larger than that 
corresponding to the disapearence of the LCP pocket. On the other hand 
for tip distances smaller than 6 fm the emission of the $\alpha$ is
strongly hindered by a thick barrier even for a rather high zero energy 
$E_{\alpha}^0\ge 3$MeV (see Fig.3).

If we choose $d$ = 8 fm for the example considered in Figure 1, 
then we get for the total kinetic energy of
the two main fragments $TKE$=46.21 MeV which is much larger than the 
corresponding kinetic energy in the spontaneous fission. For 
$d$ = 6 fm the kinetic energy will drop to $TKE$=28.78 MeV.  
Repeating this calculation for other mass splittings we conclude that 
the kinetic energy of the main fragments is ranging in the broad interval 
25 - 50 MeV, but as we shall see bellow it is correlated to the kinetic 
energy of the emitted alfa particle through the tip distance.

We are left now with the determination of the LCP geometrical and
dynamical initial characteristics. For that we invoke a receipt proposed 
by Boneh et al.\cite{BFN67} which consider as a possible choice for the
LCP position, the point of minimum potential energy (the saddle point of the 
potential energy surface). If the heavy-fragments would have to interact 
with the LCP via point-like Coulomb forces, this electrostatic saddle point 
would be determined by $Z_H/R_{\alpha H}^2=Z_L/R_{\alpha L}^2$, where 
$R_{\alpha i}(i=L,H)$ is the distance between the LCP and the main fragment 
$i$. It is readily seen from Figure 2a,b that in the case of our deformation 
dependent cluster model, where the nuclear forces are introduced via the 
M3Y potential, this saddle point corresponds to the position where the 
combined Coulomb and nuclear forces exerted by the heavy fragments on the 
LCP cancel each other and the potential surface will have a relative minima 
at this point.
To establish more precisely the location of this {\em electro-nuclear} 
saddle point, 
we use the multipolar decomposition for the M3Y potential\cite{CL92}, and 
the above  condition translates to 
\beq
\sum_{\lam}
\frac{\partial V_{~\lam 0 \lam}^{~0 0 0}(R_{\alpha H})}
{\partial R_{\alpha H}} = 
\sum_{\lam}
\frac{\partial V_{~\lam 0 \lam}^{~0 0 0}(R_{\alpha L})}
{\partial R_{\alpha L}} 
\eeq
which is a generalization of the point-like Coulomb equilibrum condition.
In the laboratory frame of reference, we choose the $z$-axis as the initial 
fissioning axis of the two heavier fragments, with the origin at the tip 
of the left (heavy) fragment. Then the location of the electrostatic saddle 
point is given analitically by the formula
\beq
z_{\alpha}(d) = \frac{d+a_L+a_H}{1+\sqrt{Z_L/Z_H}} - a_L
\eeq        
where $a_{i}~(i=L,H)$ are the major axes of the quadrupole deformed main 
fragments.  
For the pair considered in Fig.1-3, $z_{\alpha}(d)\approx$0.58$d$ using 
the point-like Coulomb forces and 0.51$d$ in our model where nuclear
forces are included to. 

As we already noted above the potential energy of the LCP positioned 
at the electro-nuclear saddle point will have a minimum in the 
$y$-direction. It is clear that the LCP can have no component
of its velocity along the $x$-axis since this would 
result in a possible absorbtion of the LCP by the deep potential wells of 
the two heavier fragments instead of being emitted
\footnote{In the case we would employ forces with repulsive nuclear core 
the LCP will be once again prevented to move in the $z$-direction.}. 
The only possibility for the LCP to {\em survive} the descent of the decaying
system from scission to the release point is to have a momenta directed 
only along the $y$-axis.  As can be infered from Figure 2 the locus of 
the saddle point is on the bottom of the potential well. Taking 
sections of the potential surface  along the $y$-axis at $z$ corresponding 
to the saddle point, the resulting potential slice will look similar 
to a one-dimensional harmonic potential well (see Figure 3).  When the tip
distance increases, the well becomes more and more shallower untill it 
vanishes completely. 
Following an idea from \cite{RRD82} we will approximate the potential 
$V_{LCP}$ with an harmonic potential in the $y$-direction, centered at 
the saddle-point
\beq
V_{LCP}\approx V_{saddle} + {1\over 2}C y^{2}
\eeq
where $V_{saddle}=V_{LCP}(z=z_{saddle},y=0)$ and
$C=\left.\frac{\partial^2 V_{LCP}}{\partial y^2}\right|_{y=0}$ is the 
stifness. It can be shown after some algebra that the elastic constant 
value is given by the expression
\beq
C=\sum_{i=L,H}{1\over R_{\alpha i}^0}\sum_{\lam\ge 0}
\left ( 
\left.  \frac{\partial V_{\lam 0 \lam}(R_{\alpha i})}
{\partial R_{\alpha i}}\right |_{0} -\frac{\lam(\lam+1)}{2}
\frac{V_{\lam 0 \lam}(R_{\alpha i})}{R_{\alpha i}}\right )
\eeq  
where $R_{\alpha i}$ is the distance from the fragment $i$ to the LCP
($\alpha$) on the $z$-axis :
\beq
R_{\alpha L}^{0}=\frac{D}{1+\sqrt{Z_L\over Z_H}},
R_{\alpha H}^{0}=\frac{D}{1+\sqrt{Z_H\over Z_L}},
\eeq
where $D$ is the inter-fragment distance. From here we get an 
estimation for the initial kinetic energy of the LCP supposing that 
it can be identified with the zero-energy in the {\em harmonic} potential 
well, i.e.
\beq
E_{LCP}={1\over 2}\hbar\sqrt{C\over m_{LCP}}
\eeq
Consequently, a degree of uncertainty in the initial kinetic energy 
occurs also for the LCP. For increasing tip distance the kinetic energy
of the LCP decreases. One might suppose that in the range 
6 - 8 fm, for the tip distance, the  LCP has the possibility to 
escape by tunneling or by the dissapearence of the barrier.
Further the velocity corresponding to this kinetic energy, 
$v_{\alpha}=\sqrt{2 E_{LCP}\over m_{LCP}}$ will have a nonzero 
component only with respect to the $y$-axis, according to the above 
discussion.
 
\section{Trajectory Equations}

In order to write down the equations of motion we have to establish 
the geometry of the system not only at the beginning but also 
long after the release of the LCP. 
The forces being central, and the initial velocities are in-plane 
the problem is simplified by a two-dimensional approximation. There will
be required six coordinates and six velocities, which are governed by 
a system of twelve first order ordinary differential equations.  
In Fig. 4, we see the three fragments and the 
forces acting between them, just after the release of the LCP. 
Contrary to other works we take into account the forces exerted 
by the LCP on the fragments. After deriving the initial conditions in 
the previous section, taking into account the nuclear forces in the 
calculation of the barriers, we proceed now to the calculation of the 
trajectories by considering only the Coulomb forces. Since the kinetic
energies of the fragments are rather high, this approximation is good even  
in the point-charge approximation. 

In previous papers \cite{Flor2,SCM98,SMC98} we used a double 
folding potential for the heavy-ion interaction. Presently we shall
consider only the Coulomb part of this interaction, between two ions, i.e.
\beq
V_C(\bd R) = \int d \bd{r}_1 \int d \bd{r}_2
\frac{\rho_1(\bd{r}_1)\rho_2(\bd{r}_2)}{|\bd{r}_1+\bd{R}-\bd{r}_2|}
\eeq 
where $\rho_{1(2)}(\bd{r})$ are the charge ground-state one-body densities 
of the fragments. The one-body densities are taken as Fermi distributions 
in the intrinsic frame for axial-symmetric nuclei
\beq
\rho(\bd{r}) = \frac{\rho_0}{1+e^{\frac{r-R(\theta)}{a}}}
\eeq
with $R(\theta)=R_{0}(1+\sum_{\lam\ge 2}\beta_{\lam}Y_{\lam 0}(\theta,0))$.
In what follows we consider that the symmetry axes of the fragments 
are lying in the same plane. Using the formalism presented in 
\cite{CL92}, the interaction between two heavy ions with orientation 
$\omega_{1},\omega_{2}$ of their intrinsic symmetry axes with respect to 
the fixed frame, reads : 
\beq 
V(\bd{R}_{12}) = \sum_{\lambda_1,\lambda_2,\lambda_3,\mu}
V_{\lam_1\lam_2\lam_3}^{\mu -\mu 0}(R_{12})
P_{\lam_1}^{\mu}(\cos\omega_{1})P_{\lam_2}^{-\mu}(\cos\omega_{2})
P_{\lam_3}(\cos\Omega_{12})
\eeq
where
\beq
V_{\lam_1\lam_2\lam_3}^{\mu -\mu 0}(R_{12})=
{2\over \pi}i^{\lambda_1-\lambda_2-\lambda_3}{\hat \lambda_1}
{\hat \lambda_2} \cleb{\lambda_1}{\lambda_2}{\lambda_3}{~0}{~0}{~0}
\cleb{\lambda_1}{\lambda_2}{\lambda_3}{\mu}{-\mu}{~0}
F_{\lam_1\lam_2\lam_3}(R_{12})
\eeq
with $P_{\lam_1}^{\mu}(\cos\omega_{1})$ and 
$P_{\lam_2}^{-\mu}(\cos\omega_{2})$ being the associated Legendre 
polynomials which describe the relative orientation of the two fragments 
whereas $P_{\lam_3}(\cos\Omega_{12})$ describes the orientation of the axis 
joining the two nuclei with respect to the laboratory frame.
In the present study the LCP is spherical and thus the interaction between 
the  LCP and one heavy fragment $i$(=$L,H$) will get a simplified form 
\beq
V(\bd{R}_{\alpha i}) = \sum_{\lam}
V_{~\lam 0 \lam}^{~000}(R_{\alpha i})P_{\lam}(\cos\theta_i)
\eeq
The following approximation can be applied for the two heavy fragments : 
Since their relative orientation does not change significantly at the 
beginning of the quasiclassical motion, one can neglect the relative 
orientation of the heavy fragments: 
\beq
V(\bd{R}_{LH}) = \sum_{\lam_1\lam_2\lam_3}
V_{~\lam_1 \lam_2 \lam_3}^{~0~0~0}(R_{LH})P_{\lam_3}(\cos\theta_{LH})
\eeq

The force acting between a pair of fragments can be written: 
\beq
\bd{F}_{ij} = -\bd{\nabla} V(\bd{R}_{ij})
\eeq
The force acting between the two heavy fragments is given by:
{\small
\beqa
&\bd{F}_{LH}& = -\bd{e}_x\sum_{\lam_1\lam_2\lam_3}
\left (
\frac{\partial V_{~\lam_1\lam_2\lam_3}^{~0~0~0}(R_{LH})}{\partial R_{LH}}
P_{\lam_3}(\cos\phi)\cos\phi -
\frac{V_{~\lam_1\lam_2\lam_3}^{~0~0~0}(R_{LH})}{R_{LH}}
P_{\lam_3}^{1}(\cos\phi)\sin\phi \right )
\nonumber\\
&-&~\bd{e}_y\sum_{\lam_1\lam_2\lam_3}
\left (
\frac{\partial V_{~\lam_1\lam_2\lam_3}^{~0~0~0}(R_{LH})}{\partial R_{LH}}
P_{\lam_3}(\cos\phi)\sin\phi +
\frac{V_{~\lam_1\lam_2\lam_3}^{~0~0~0}(R_{LH})}{R_{LH}}
P_{\lam_3}^{1}(\cos\phi)\cos\phi \right )
\eeqa
}
whereas the forces exerted by the fragments on the LCP read: 
{\small
\beqa
&\bd{F}_{H\alpha}& = -\bd{e}_x\left ( \sum_{\lam\ge 0}
\frac{\partial V_{~\lam~0~\lam}^{~0~0~0}(R_{\alpha H})}
     {\partial R_{\alpha H}}P_{\lam}(\cos\psi_1)\cos\psi_1 -
\sum_{\lam\ge 2}
     \frac{ V_{~\lam~0~\lam}^{~0~0~0}(R_{\alpha H})} { R_{\alpha H}}
     P_{\lam}^{1}(\cos\psi_1)\sin\psi_1\right )
\nonumber\\
& - &~\bd{e}_y \left (
\sum_{\lam\ge 0}
\frac{\partial V_{~\lam~0~\lam}^{~0~0~0}(R_{\alpha H})}
     {\partial R_{\alpha H}}P_{\lam}(\cos\psi_1)\sin\psi_1 +
\sum_{\lam\ge 2}\frac{ V_{~\lam~0~\lam}^{~0~0~0}(R_{\alpha H})} { 
R_{\alpha H}}
P_{\lam}^{1}(\cos\psi_1)\cos\psi_1 \right )
\eeqa

\beqa
&\bd{F}_{L\alpha}& = \bd{e}_x\left ( \sum_{\lam\ge 0}
\frac{\partial V_{~\lam~0~\lam}^{~0~0~0}(R_{\alpha L})}
     {\partial R_{\alpha L}}P_{\lam}(\cos\psi_2)\cos\psi_2 -
\sum_{\lam\ge 2}
     \frac{ V_{~\lam~0~\lam}^{~0~0~0}(R_{\alpha L})} { R_{\alpha L}}
     P_{\lam}^{1}(\cos\psi_2)\sin\psi_2\right )
\nonumber\\
& - &~\bd{e}_y \left (
\sum_{\lam\ge 0}
\frac{\partial V_{~\lam~0~\lam}^{~0~0~0}(R_{\alpha L})}
     {\partial R_{\alpha L}}P_{\lam}(\cos\psi_2)\sin\psi_2 +
\sum_{\lam\ge 2}\frac{ V_{~\lam~0~\lam}^{~0~0~0}(R_{\alpha L})} { 
R_{\alpha L}}
P_{\lam}^{1}(\cos\psi_2)\cos\psi_2 \right )
\eeqa
}
The equations of motion of the three nuclei are:
\beqa
M_{L}\ddot{\bd{r}}_{L}& = &~\bd{F}_{LH}-\bd{F}_{L\alpha}
\nonumber\\
M_{H}\ddot{\bd{r}}_{H}& = &-\bd{F}_{LH}-\bd{F}_{H\alpha}
\nonumber\\
m_{\alpha}\ddot{\bd{r}}_{\alpha} & = & \bd{F}_{L\alpha}+\bd{F}_{H\alpha}
\eeqa

Here we assumed that the two heavy fragments have the same multipolarity 
in deformations. In this paper we consider only quadrupole 
deformations.

The above system was solved numerically employing the {\tt lsoda}
package for ordinary differential equations, with automatic method 
switching for stiff and nonstiff problems \cite{linda83}.

In Figure 5 a, b, c we presented the trajectory 
of the three fragments for the two extreme initial conditions 
(with high and with low kinetic energies of the heavier fragments)
in a sequence of 10 time steps. The time scale is divided into
increments of $\Delta t$ = 1.8$\times$ 10$^{-22}$.
In figure 5a we display the trajectories of one of the most asymmetric 
splittings, recorded in experiment, i.e. $^{152}$Nd + $^{92}$Kr. 
Since in this case the $\alpha$ feels a stronger repulsion from the 
heavy fragment, it will be deflected at a larger angle in the direction of the
light fragment. 
In the case of the splittting $^{144}$Xe + $^{104}$Nd 
this deflection will be less pronounced (Fig.5b) and for the more 
equilibrated splitting, i.e. $^{132}$Nd + $^{116}$Pd, the 
$\alpha$ will be only slightly deflected (Figure 5c). 
We thus observe that in all the cases the $\alpha$-particle is deflected 
in the direction of the light fragment, but with a larger 
angle when the initial kinetic energy of the heavier fragments is higher.
This fact should be attributed to the low energy of the $\alpha$ ($\leq$
1MeV) which makes it to feel for a longer time the repulsion coming from 
the heavy fragment. 
In Table I we present the final kinetic energy $E_{\alpha}^f$ and 
the asymptotic angle $\theta_{\alpha}^f$ for the three splittings mentioned 
above  when we employ point-like and size dependent forces.
In all cases we observe the decreasing of $E_{\alpha}^f$ with increasing 
tip distance $d$. The mean of these two energies is not far from the value 
of $\langle E_{\alpha}^f\rangle$ = 16.0 $\pm$0.2 MeV which is the most 
probable $\alpha$-particle energy predicted by the trajectory calculation 
for the {\em hot} fission.
Thus, the phenomenon of $\alpha$-particle energy {\em amplification} in the
cold fission seems to follow the same pattern like in normal fission.
This effect should be attributed solely to the predominant effect 
of the Coulomb field and less to deformation or finite size effects.  
It should also be remarked the near constancy of the final LCP kinetic
energy for different mass splitings at the same tip distance, a fact already
remarked long time ago in spontaneous fission\cite{Halp71}.
In what concerns the angles at which $\alpha$-particles are emitted their 
dependence on the mass splitting is obvious. Deviations from the axis 
perpendicular to the fission axis increase with the mass ratio.   
The difference observed between the two sets of data points to an 
important influence of the geometrical factors, which however does not 
alter the general trends of the process.

Naturally one might next ask if the experimental status of the
problem allows the comparison with the results presented in this paper.
Up to now there are no special data on cold fission
available since from the expermental side it would mean to
set  a trigger on neutronless events, which is very difficult to attain in 
practice. There are available data on the alpha (and other particles) 
spectra, as a function of the total excitation energy, reaching  TXE = 10 MeV 
within the experimental accuracy performed by the Darmstadt group with 
the DIOGENES setup, and in a more recent work at the MPI Heidelberg
\cite{Mut98}.
These data does not contain special effects in the alpha spectra,
when the cold fission regime is approached, except that the mean energy
increases nearly linearly with decreasing TXE. This would mean that if 
the linear dependency would be extrapolated to TXE =  0 MeV, i.e. when the
scission configuration tends to become compact, like in our model, 
the average kinetic energy of the $\alpha$ will approach the value
17.5 MeV\cite{Mut98}.
According to the calculations presented in this work a range between 12 MeV
to 20 MeV should be expected for the final kinetic energy if we consider
that the $\alpha$ particle occupies the lowest states in the pocket formed
from the interaction with the two heavier fragments.
Therefore the experiment doesn't show a distinctive $\alpha$ kinetic
energy distribution for cold fission, a fact which is in agreement with
the calculations we presented in this paper. In order to establish more  
precisely $\langle E_{\alpha}\rangle$ we should carry out Monte-Carlo
calculations. The fact that the experimental value is slightly higher than
in hot fission (15.9 MeV) is a sign that the $\alpha$ is emitted
earlier in cold fission, according to the uncertainty relation for energy  
$\Delta E\cdot\Delta t\approx\hbar$.

\section{Final remarks}

We presented a receipt to determine the initial conditions for trajectory 
calculations in the ternary cold fission of $^{252}$Cf. Compared to the 
case when the fragments are emitted with high excitation energy, the 
initial conditions are restricted to a narrower range of values as a 
consequence of the pecularities of the process, mainly the compact shape 
of the fragments. 
 
In our model the $\alpha$-particle cannot be emitted at a tip distance 
larger than 8 fm, because as we showed for such a distance the 
$\alpha$ particle is no longer under the influence of the attractive 
nuclear forces, and on the other hand we disregard tip distances smaller
than 6 fm because the LCP wavepacket filling the lowest state in the potential
well has a small probability to tunnel through the thick transversal 
barrier.   

The location of the LCP was fixed at the electro-nuclear saddle point, which
is also model dependent. Due to the finite size and the deformations of the 
fragments this location will be shifted with respect to the location of the
electrostatic saddle point, and consequently the outcome of the 
trajectory calculation will be altered to a certain extent. 

The initial kinetic energy of the LCP was considered to coincide with the
lowest level occupied in a one-dimensional harmonic potential well oriented
perependiculary to the fission axis. This energy is decreasing with the 
tip distance.

As have been pointed by Halpern \cite{Halp71}, there is no reason to 
believe that the third-particle ejection rates should be independent 
of the initial angular momentum. In our case the spin of the parent
nucleus ($^{252}$Cf) being zero the angular momentum imparted to the
fragments and their relative angular momentum is mainly due to the 
creation of a  molecular configuration at the scission point\cite{Mis97}. 
In the model presented in this paper we didn't took into account the influence 
of collective molecular excitations, like bending or wriggling, nor the
torques exerted between the fragments during the quasiclassical motion. 
The inclusions of these suplementary degrees of freedom  
could alter the initial configuration. This would be an interesting 
topic for a future investigation of the scission configuration in ternary 
cold fission. Moreover the evolution from scission to the release point 
of the LCP should be done in a dynamical way, i.e. writing equations of
motion not only for the translational and rotational degrees of motion but 
also for the dynamical change of deformation, because one might suppose that 
even if the {\em trinuclear} system is almost cold the small excitation
energy present in the reaction will induce a $\beta$-polarization of
the fragments \cite{Avr95}.

\section{Acknowledgements}

I would like to express my gratitude to prof.Ter-Akopian and
prof.F.G\"onnenwein for fruitfull discussions and encouragements during
the completion of this work and to M.Rizea and dr.F.C\^ arstoiu for their 
kindness in offering me some routines used in the calculations. 
The informations on the experimental status of the alpha accompanied
cold ternary fission provided by Prof.Mutterer and Dr.Kopachi is
acknowledged. I am also very indebted to prof.Halpern who carefully
readed the manuscript and expressed critical remarks. 

\begin{table}
\caption{The tip distance $d$, the initial kinetic energy, 
$E_{\alpha}^{0}$, the final kinetic energy $E_{\alpha}^{f}$ and the 
asymptotic angle $\theta_{\alpha}^{f}$ of the $\alpha$, with point-like 
and with finite size Coulomb forces}
\label{table:a}
\begin{tabular}{c c c c c c c}
Splitting & $d (fm)$ & $E_{\alpha}^0$ (MeV) & 
$E_{\alpha}^f$ (MeV)& 
$\theta_{\alpha}^f$& 
$E_{\alpha}^f$ (MeV)& 
$\theta_{\alpha}^f$\\ 
& & & \multicolumn{2}{c} {Point-like forces} &
\multicolumn{2}{c} {Finite-size  forces}\\ 
\hline
 & 6 & 2.71 & 20.10& 80.83 & 21.36 & 77.86\\
$^{92}$Kr+$^{156}$Nd & 7 & 1.72 & 15.84 & 78.77& 16.91& 75.70 \\
 & 8 & 0.85 & 11.25 & 75.42& 12.10 & 73.14\\  
\hline
 & 6 & 2.68 &19.87 & 85.03 & 20.96&83.29\\
$^{104}$Mo+$^{144}$Xe & 7 & 1.70 &15.57 &83.77 &16.44 &82.08\\
 & 8 &0.82 &10.82 &81.63 &11.47 &80.32\\
\hline
 & 6 & 2.85&20.29 &88.17 &20.86 &86.99\\
$^{116}$Pd+$^{132}$Sn & 7 &1.84& 15.86 &87.69 &16.31 &86.27\\
 & 8 &0.96 &11.53 &85.68 &11.19 &86.93\\
\end{tabular}
\label{table:1}
\end{table}

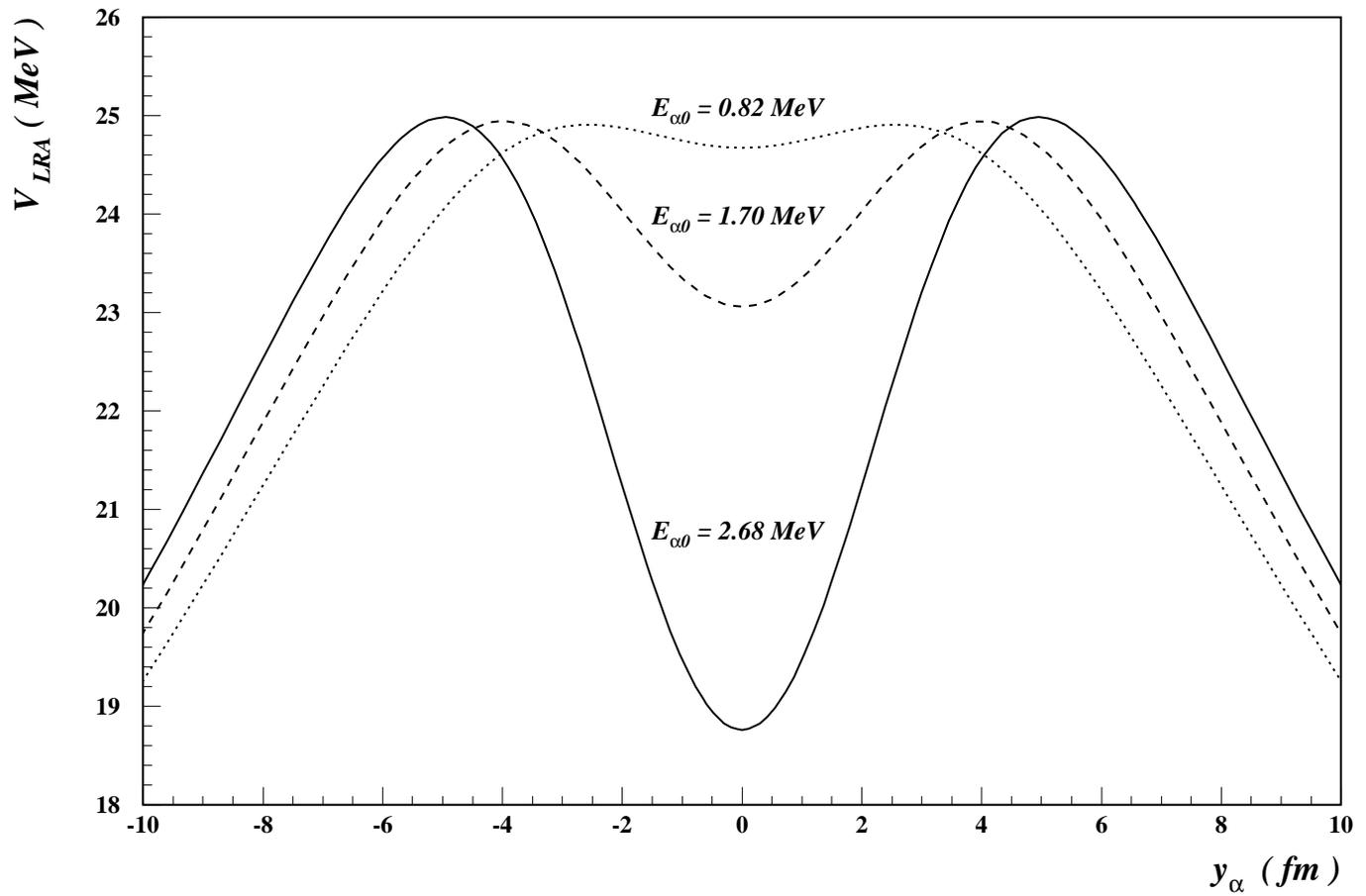
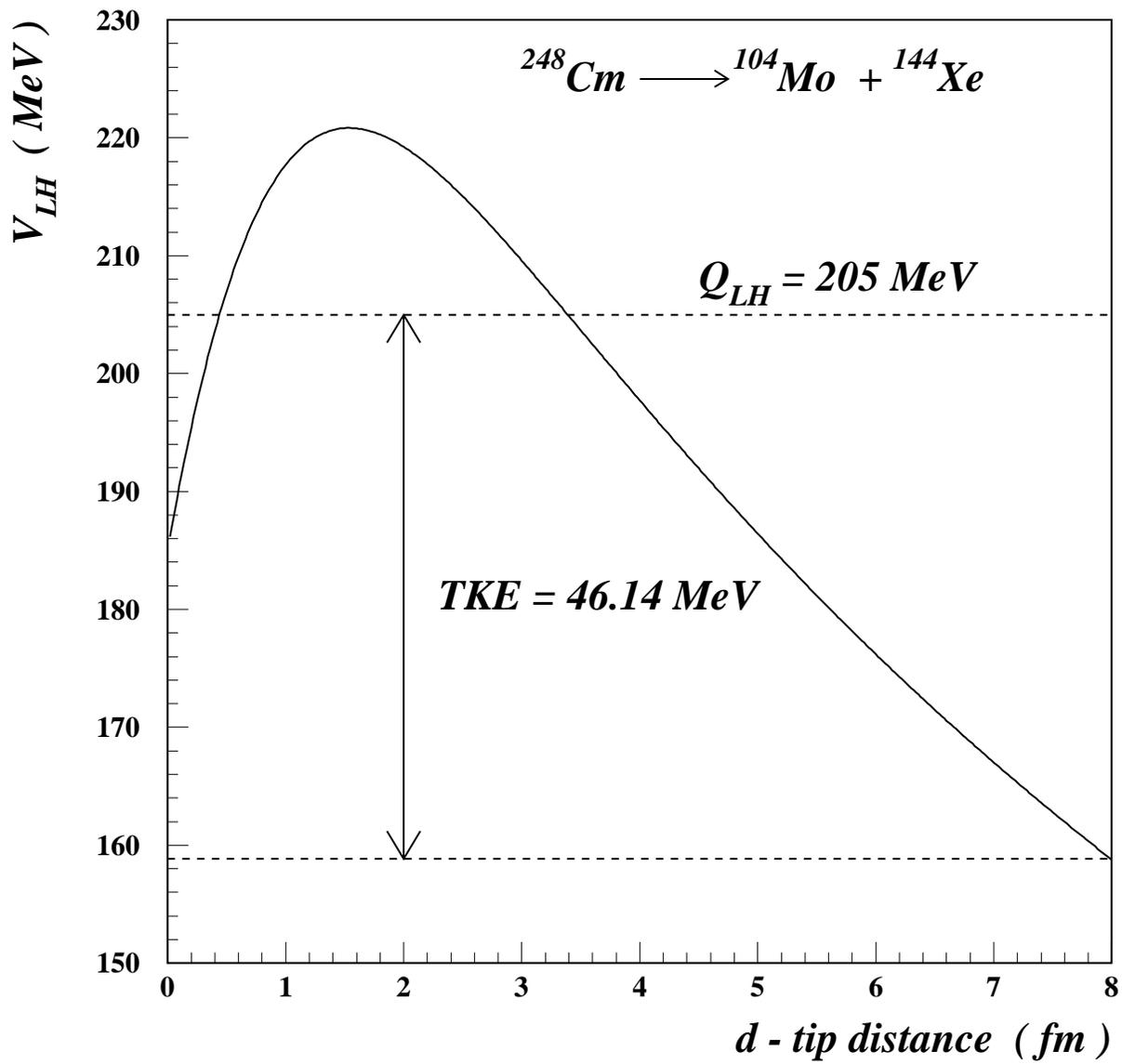
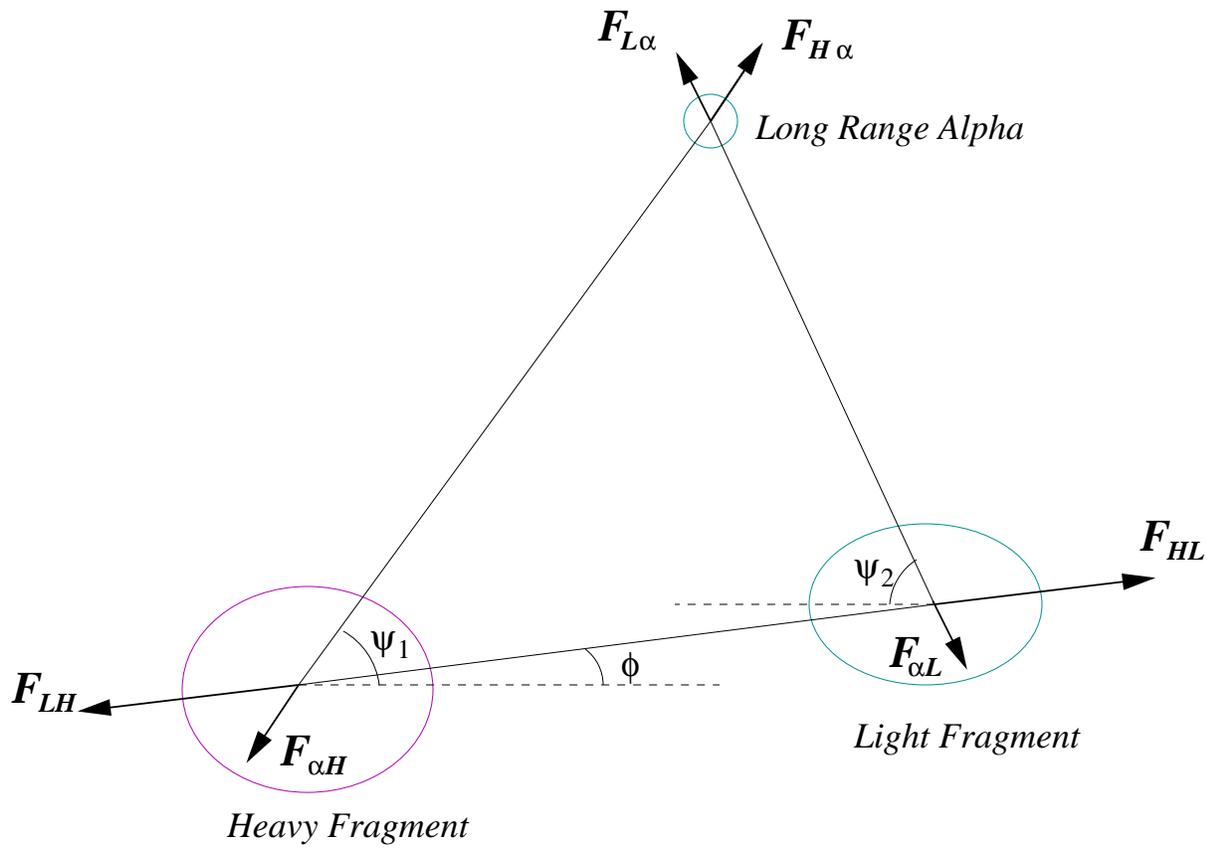
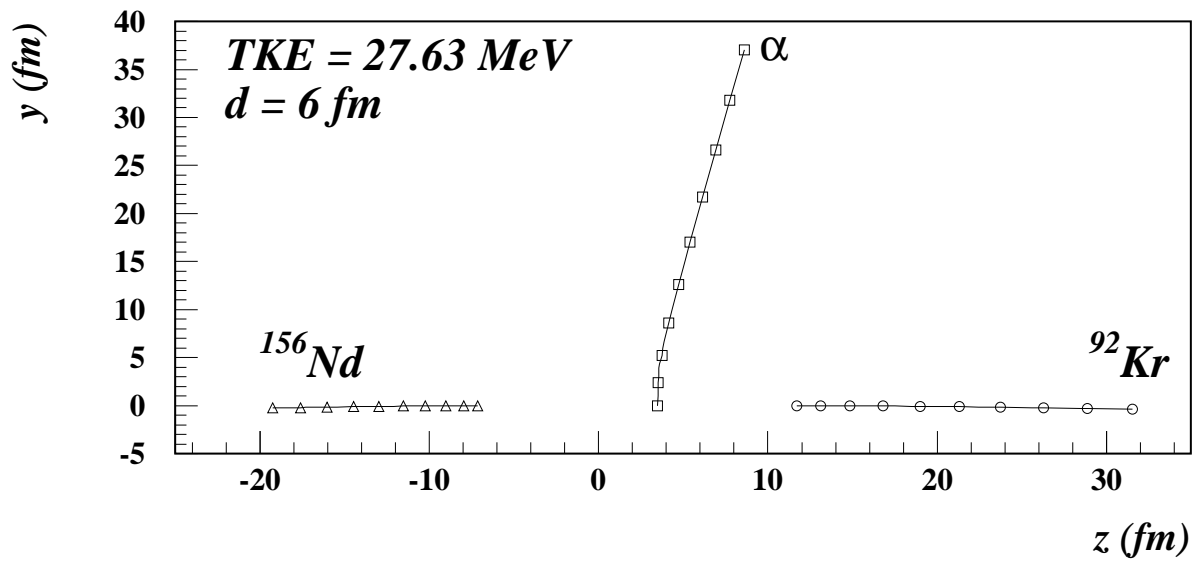
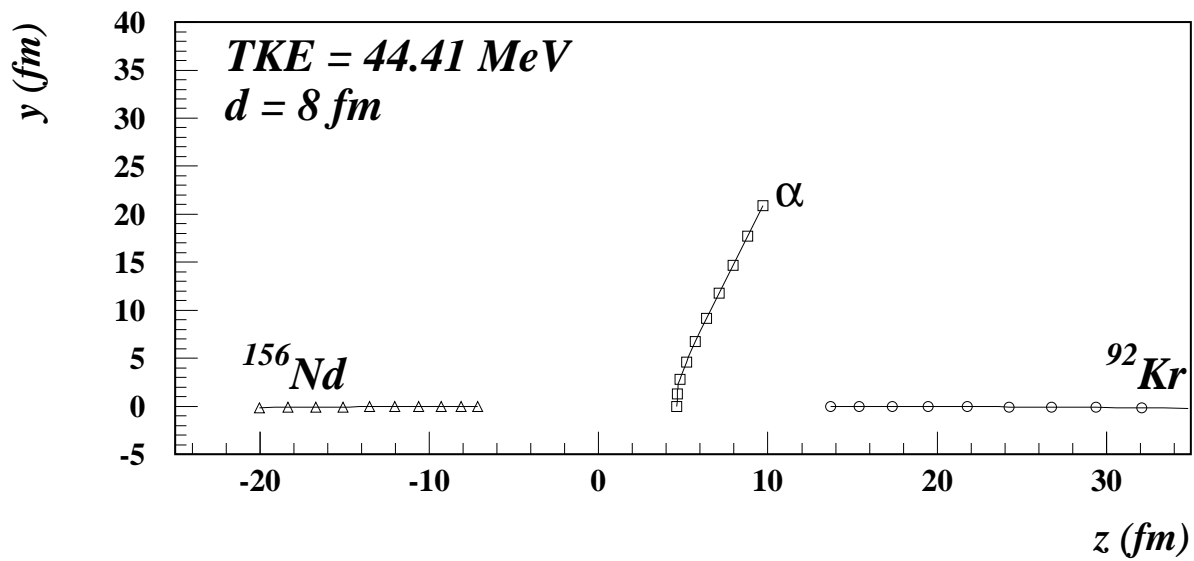
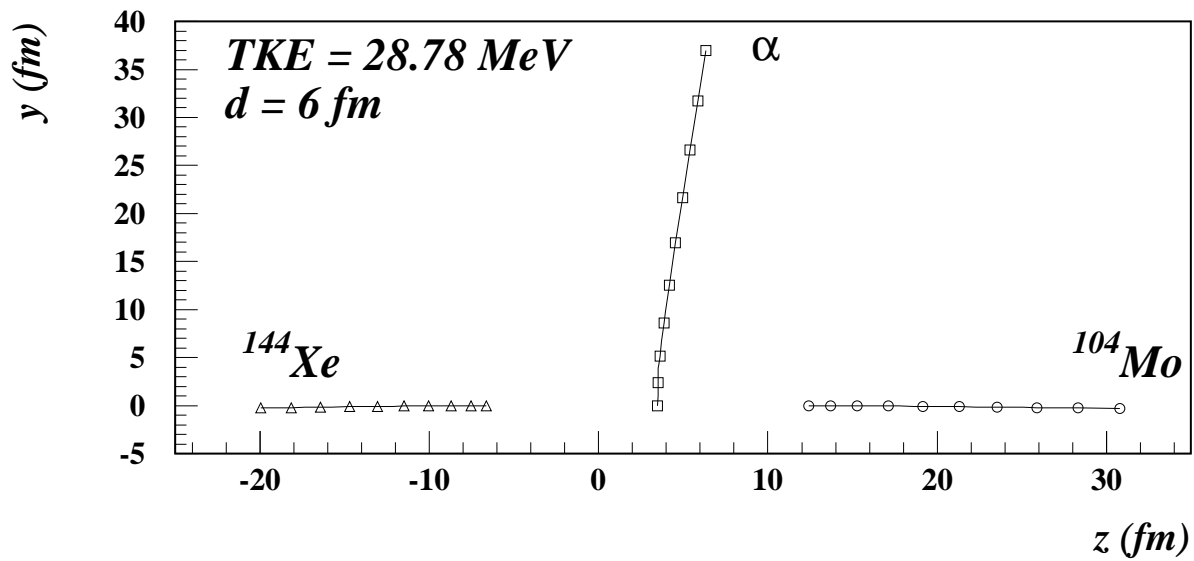
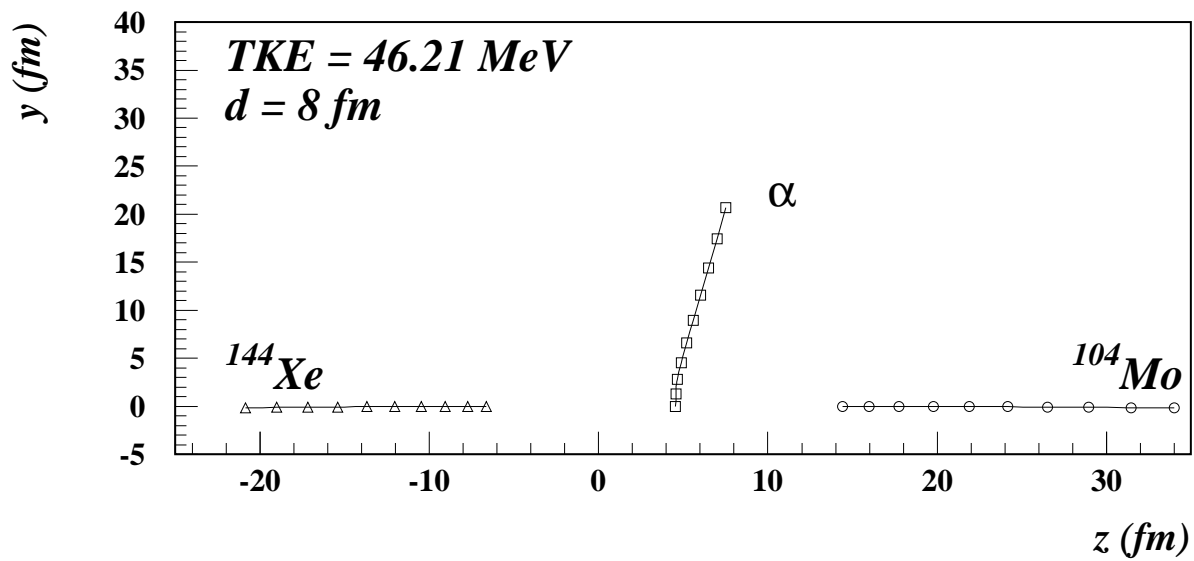
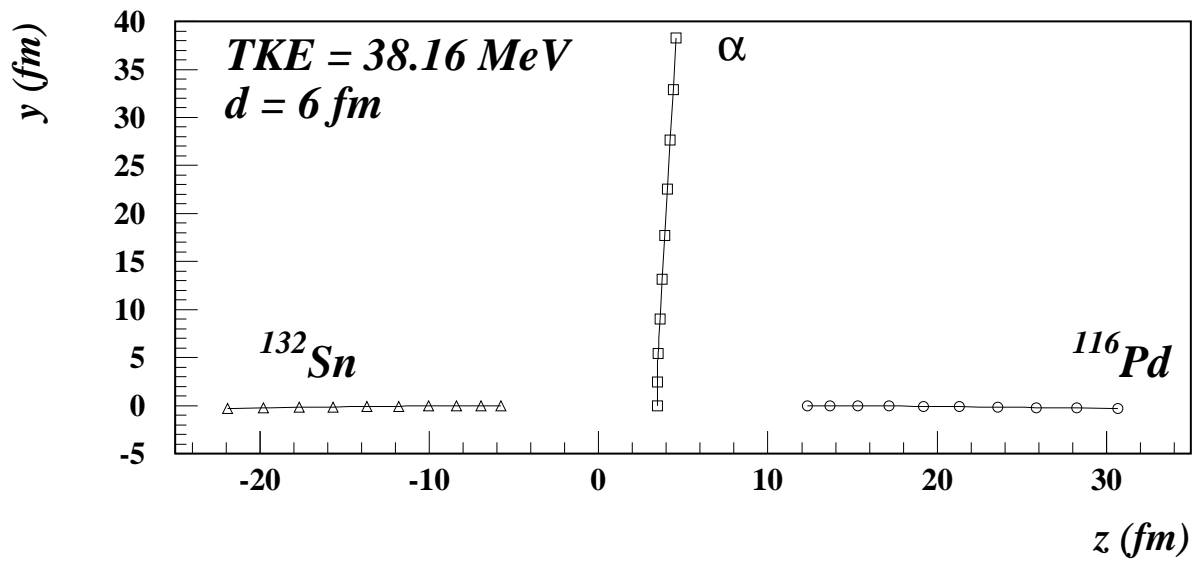
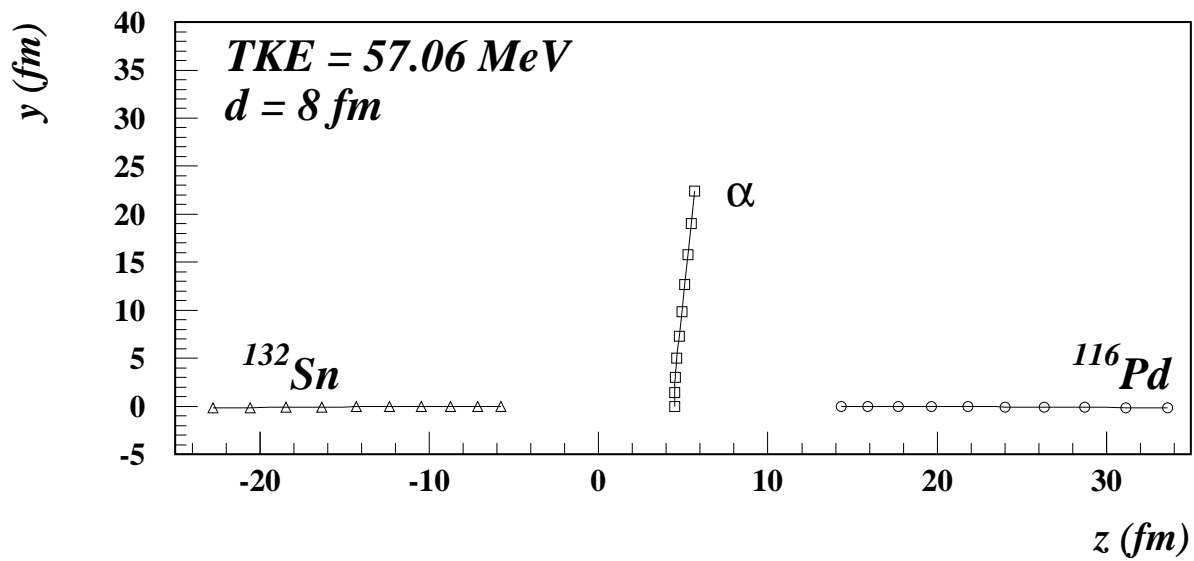
\end{document}